# Water Adsorption in Soft and Heterogeneous Nanopores

François-Xavier Coudert[a,*]

[a] Chimie ParisTech, PSL Research University, CNRS, Institut de Recherche de Chimie, Paris, 75005 Paris, France

* fx.coudert@chimieparistech.psl.eu ; https://www.coudert.name

**CONSPECTUS**

Liquids under confinement differ in behavior from their bulk counterparts and can acquire properties that are specific to the confined phase and linked to the nature and structure of the host matrix. While confined liquid water is not a new topic of research, the past few years have seen a series of intriguing novel features for water inside nanoscale pores. These unusual properties arise from the very specific nature of nanoporous materials: termed "soft porous crystals", they combine large-scale flexibility with a heterogeneous internal surface. This creates a rich diversity of behavior for the adsorbed water, and the combination of different experimental characterization techniques along with computational chemistry at various scales is necessary to understand the phenomena observed, and their microscopic origins. The range of systems of interest span the whole chemical range, from the inorganic (zeolites, imogolites)[2] to the organic (microporous carbons, graphene and its derivatives) and even encompass the hybrid organic–inorganic systems (such as metal–organic frameworks).

The combination of large scale flexibility with the strong physisorption (or even chemisorption) of water can lead to unusual properties (belonging to the "metamaterials" category) and to novel phenomena. One striking example is the recent elucidation of the mechanism of negative hydration expansion in $ZrW_2O_8$, by which adsorption of ~10 wt% of water in the inorganic nonporous framework leads to large shrinkage of its volume. Another eye-catching case is the occurrence of multiple water adsorption-drive structural transitions in the MIL-53 family of materials: the specific interactions between water guest molecules and the host framework create a behavior that has not been observed with any other adsorbate. Both are counter-intuitive phenomena, that have been elucidated by a combination of experimental *in situ* techniques and molecular simulation.

Another important direction of research is the shift in the systems and phenomena studied, from physical adsorption towards studies of reactivity, hydrothermal stability, and the effect of confinement on aqueous phases more complex than pure water. There have been examples of water adsorption in highly flexible metal–organic frameworks being able to compete with the materials' coordination bonds, and therefore limiting its hydrothermal stability — while tweaking the functional groups of the same framework can lead to increased stability, while retaining the flexibility of the material. However, this additional complexity and tuneability in the macroscopic behavior can occur from changes in the confined fluid, rather than the material. Very recent studies have shown that aqueous solutions of high concentration (such as LiCl up to 20 mol L$^{-1}$), confined in flexible nanoporous materials, can have specific properties different from the pure water, and not entirely explained by osmotic effects. There, the strong ordering of the confined electrolyte competes with the structural flexibility of the framework to create an entirely new behavior for the {host, guest} system.

**Introduction**

The last two decades have seen a rapid increase in the number of new crystalline materials reported, both dense and porous, based on molecular or supramolecular framework architectures. Maybe the most emblematic family of these, metal–organic frameworks (MOFs), are nanoporous solids formed by a hybrid organic–inorganic framework, which is in most cases (but not necessarily) crystalline. Other families of materials with framework-based architectures, such as covalent organic frameworks (COFs), supramolecular organic frameworks (SOFs), and porous molecular solids, have also been synthesized, characterized and are seeing a swift development. Compared to dense inorganic solids or porous materials such as zeolites, this "new generation" of nanoporous solids relies on three-dimensional assembly through relatively weak interactions (metal–ligand coordination, π–π stacking, hydrogen bonds, etc.), giving them a high intrinsic structural flexibility.[1,2] This is coupled with an important molecular complexity of the materials' building blocks (especially the organic parts), which provides great versatility in the possible chemical compositions and gives rise to a large number of internal degrees of freedom.[3] The name of "soft porous crystals" has been proposed to designate such materials,[4] however they are also often referred to as stimuli-responsive, multifunctional, smart, or dynamic materials.[5,6]

The flexibility in soft porous crystals does not manifest itself only in mechanical terms, i.e., a high degree of "compliance" or "softness" under compression. More generally, these flexible frameworks demonstrate an impressive capacity to respond to external stimuli by large-scale changes in their structure, which consequently triggers significant changes in their physical or chemical properties. This coupling of two or more properties (temperature, pressure, electric field, magnetic field, pH, ionic strength, host molecules, etc.) is multifaceted, and the materials behaving as such are structurally and chemically very diverse. They have been widely studied, both for the beauty and sheer complexity of their architectures, for the difficulty of their characterization and the understanding of these dynamic phenomena, but also for applications as nanosensors,[7,8] nanoactuators or nanoswitches,[9] for gas or liquid phase separation,[10,11] or for the encapsulation of diverse active ingredients.[12,13]

Over time, an increasing number of couplings between physical or chemical phenomena have been brought to light — and there are certainly still others to be discovered. However, such studies are rendered difficult by the complexity of the phenomena involved and the need to characterize them experimentally by carrying out *in situ* measurements of several properties simultaneously. Advanced experimental characterization techniques, along with computational chemistry methods, are playing a leading role in this quest.

As introduced above, the impact of temperature, pressure, and guest adsorption on framework materials in general, and soft porous crystals in particular, have been widely studied. However, in most of these cases, the adsorbate molecules considered were in the gas phase — i.e., the adsorption of the fluid occurs below the saturation pressure, which corresponds to a situation with relatively weak interactions with the nanoporous materials — a regime of weak physisorption. The situation is different when we consider the adsorption of liquids, and the properties of those liquids confined into spaces of nanometric dimensions. This is particularly true for the adsorption of water, and the behavior of liquid water confined in nanopores, due to the important and very directional water–water interaction, and the highly structured nature of water in the liquid phase.[14,15] Confined phases of water often demonstrate very specific properties, widely different from the bulk liquid, weather the confinement takes place in hydrophilic porous materials (which feature strong physisorption, or even in some cases, chemisorption) or in hydrophobic pores (where the water–sorbent interactions are weaker than water–water interactions).

The combination of the large scale flexibility and heterogeneous internal surface of soft porous crystals, with the strong interactions of water creates a rich diversity of behavior for the adsorbed water, and can lead to unusual properties (belonging to the "metamaterials" category)[16] and to novel phenomena in the confined phase. This paper highlights some of these emerging concepts, exhibiting how they can be understand through a synergy between different experimental characterization techniques and computational chemistry at various scales.[17]

**Water in heterogeneous nanopores of MOFs**

The internal surface of the metal–organic frameworks is chemically heterogeneous, as it incorporates both organic and inorganic components, often with some polar domains in the internal pore surface (with strong affinity for water) and some aromatic entities (which tend to be hydrophobic). As water is a polar fluid, its interaction with the internal surface of MOFs can therefore be particularly complex. Understanding the adsorption of water, and the impact of confinement on the properties of adsorbed water, is an issue of both practical and fundamental interest, because water plays a major role in many industrial processes (separation, catalysis, etc.) and biological systems.[18] Moreover, specific applications have been proposed for water/MOF systems, including water harvesting from humid air in the desert[19] and low-temperature actuated adsorption-driven chillers.[20]

The behavior of water confined in the pore space of MOFs typically aims at understanding its structure, dynamics, and electronic properties, and comparing them with water in the liquid phase, or confined in other porous materials. Experimental and computation tools to achieve this goal are numerous, including *in situ* spectroscopy, X-ray diffraction, NMR, and both Monte Carlo and molecular dynamics modelling methods. It is often the combination of different such methods that allows to obtain a broad picture and full understanding of the nature and behavior of confined water. One good and recent example of this: Rieth et al. combined infrared spectroscopy and many-body molecular dynamics simulations to understand the specific hydrogen bonding structure of water in $Co_2Cl_2BTDD$ (H2BTDD = bis(1*H*-1,2,3-triazolo[4,5-b],[4′,5′-i])dibenzo[1,4]dioxin)).[21] They could show that the structure of adsorbed water evolves as a function of water loading, from one-dimensional chains of hydrogen-bonded water molecules near metal sites, into full pore filling with a characteristic hydrogen-bond network at higher humidity. The two hydration states, dictated by the MOF geometry and the MOF–water interactions, are depicted on Figure 1. This adsorption of water also has an impact on the framework, with unit cell dimensions varying by ~4.5% in constant-pressure simulations.

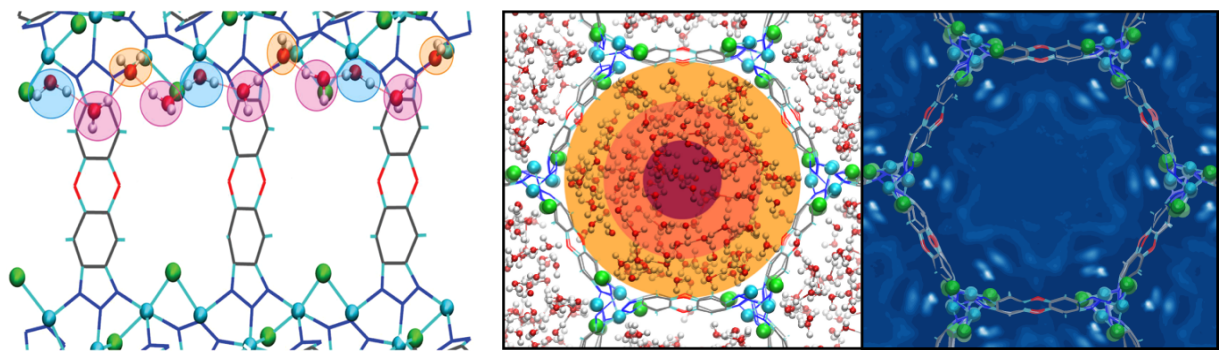

**Figure 1.** Left: One-dimensional water chain in $Co_2Cl_2BTDD$ at low loading. Right: Water at high loading, shown as a snapshot and 2D density map. Reproduced with permission from Ref. 21. Copyright The Authors, some rights reserved. Distributed under a Creative Commons Attribution License 4.0 (CC BY) https://creativecommons.org/licenses/by/4.0/

In work from my group, we have used both classical simulation as well as first-principle dynamics in order to address the question of the impact of MOF structure and internal surface chemistry on the properties of the water confined. By studying series of members of the Zeolitic Imidazolate Framework (ZIF) family of MOFs, how topology, geometry, and linker functionalization can drastically affect the adsorption of water into the pores. By classical Monte Carlo simulations, we showed that one could tune the host–guest interactions by functionalization of the organic linkers, in order to make the frameworks hydrophobic (like the parent compound ZIF-8) or hydrophilic.[22] Different types of adsorption isotherms can be measured as a response of changes in functionalization and framework topology, tuning the adsorption properties. Starting from a hydrophobic material, with water adsorption occurring in the liquid phase through a type V adsorption isotherm, upon introduction of hydrophilic functionalization we observed a gradual evolution: adsorption takes place at lower pressure, in the gas phase, and the isotherm acquires characteristics of a type I isotherm. It is also possible to construct amphiphilic frameworks, like ZIF-65(RHO), whose pore space features both hydrophobic and hydrophilic patches linked to the local effect of the very polar 2-nitroimidazolate linker.

Another illustration of the chemical heterogeneity of the internal surface of MOFs is the case of MIL-53(Cr), which is rather typical in that it features both aromatic organic linkers (which are hydrophobic) and hydrophilic inorganic parts (in this case, infinite Cr(OH) one-dimensional chains). We used first-principles molecular dynamics (MD) to investigate two hydrated phases of the flexible MOF MIL-53(Cr), which can adopts a narrow or a large pore form, depending on the water loading (see Figure 2) — although the direct simulation of the transition between the two phases, at the ab initio level, is still an open problem.[23] We found that the behavior of the confined water, its structure and dynamics, are heavily impacted by the details of the structure, resulting from a balance between the favorable water–hydroxyl interactions, and the water packing possible in the restricted pore space, compatible with water–water interactions.[24] In particular, water in the narrow-pore form is adsorbed at specific crystalline sites, while it shows a more disordered, bulk-like structure in the large-pore form. This liquid adsorbed water is, however, more strongly structured by the confining framework than bulk water, as can be seen on the water–water radial distribution functions (RDFs, Figure 2, lower panel): while the RDFs have the same shape as the bulk liquid, the intermolecular O–O and O–W peaks are higher. On the other hand, the confinement disrupts the strongly associated network of hydrogen-bonded water molecules, having an effect on the hydrogen bonding patterns (O–H⋯O angle) and tetrahedral order (O–O–O angle) distributions. Moreover, in that strongly confined state, the reorientation dynamics of the liquid water is considerably slowed down with respect to bulk water, highlighting the effect of the nanoporous framework and the water–MOF interactions.

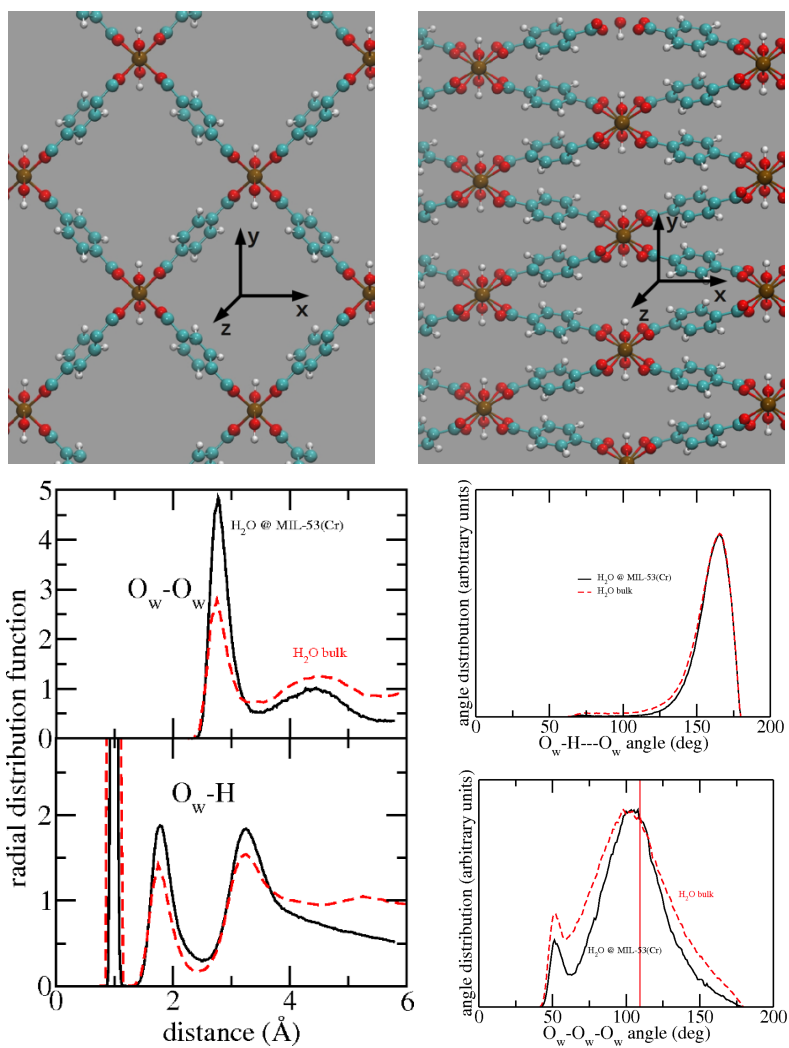

**Figure 2.** Top panel: Large-pore (left) and narrow-pore (right) structures of the MIL-53 framework. Bottom panel: Difference in structure between bulk liquid water and water confined in MIL-53(Cr) (large pore form): radial distribution functions and angle distribution functions. Reproduced with permission from Ref. 24. Copyright 2013 the PCCP Owner Societies.

**Water in soft porous crystals**

The adsorption of gases in soft porous crystals can be used as a trigger for their structural flexibility, creating adsorption-induced phase transitions. Such transitions find their microscopic origin in the stress exerted by the adsorbed molecules on the framework, the so-called adsorption stress;[25,26] this is a very general phenomenon, and therefore adsorption-induced transitions can be triggered by many different guests. This has been widely described, both experimentally and theoretically. However, the adsorption of molecules with strong or specific intermolecular interactions, such as water, can lead to specific behaviors that are not observed with other adsorbates. A few examples are presented in this section.

The first example is the impact of water adsorption on the flexibility of "breathing" MOFs of the MIL-53 family. This family of materials demonstrate bistability between an open phase (the "large pore" structure) and a phase with lower pore volume (the "narrow pore" phase). The adsorption-triggered transitions between these phases are best understood through phase diagrams, for each adsorbate, in the (temperature, guest pressure) parameter space.[27] While many simple gases such as $CO_2$, $N_2$, $O_2$, noble gases, and hydrocarbons have been published in the literature, studies on water adsorption in those "breathing" MOFs are significantly more difficult. We addressed this issue through a collaboration with experimental groups, with numerous round-trips between experimental observations, model hypotheses, predictions and their validation.[28,29] These have revealed a picture that is quite complex, in part due to the strong MOF–water interactions and the heterogeneous internal surface of the pores (as described in the previous section), but also because of their interplay with the weak interactions driving the flexibility of the host material. The resulting adsorption isotherms betray this complexity, as do the phase diagrams that rationalize them — in particular, the adsorption of water leads to the formation of new phases. For the Ga-MIL-53 material, the hydrated MOF can present two additional new phases, intermediate in structure between the "large pore" and "narrow pore" phases, which are not observed in the presence of other adsorbates (the "int" and "np_$H_2O$" phases in Figure 3). These phases, whose fingerprint are seen in X-ray powder diffractograms under certain conditions, have been identified by quantum chemistry (DFT) calculations. Their domains of existence in the phase diagram ($T$, $P$) could then be established on the basis of adsorption data (isotherms and isobars of water adsorption) interpreted by thermodynamic models. We can see on Figure 3 (left panel) that the isotherms present several steps and regimes, and would be impossible to interpret directly.

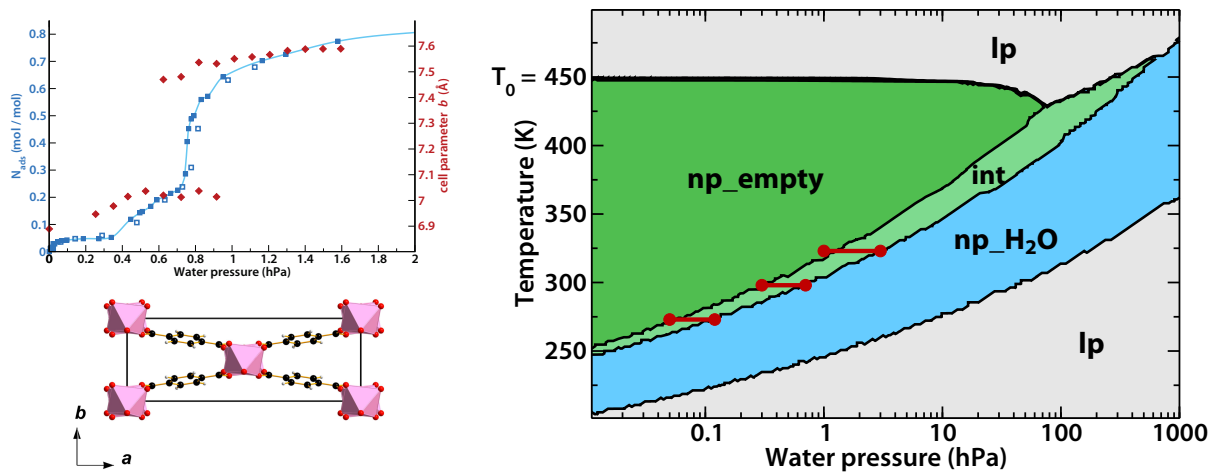

**Figure 3.** Left: Experimental water adsorption−desorption isotherm at 298 K on Ga-MIL-53 (in blue; filled symbols and line, adsorption; empty symbols, desorption), and evolution of unit cell parameter *b* under adsorption, as measured by *in situ* XRD. Right: water pressure vs temperature phase diagram of Ga-MIL-53, indicating the domains of stability of the different structures of the material. Reproduced with permission from Ref. 28. Copyright 2014 American Chemical Society.

Such studies have also shown that the occurrence of such phases is intimately linked to the minute details of the water–MOF interactions: for example, in the sibling material Al-MIL-53, the stable phase at room temperature is the large pore phase (and not the narrow pore phase, as in Ga-MIL-53), and there are no intermediate hydrated phases observed. By looking at the energetics and thermodynamics of the two materials, and the adsorption properties, computational chemistry can explain such differences in terms of the coordination chemistry and molecular orbitals of the two metals (due to the more diffuse nature of the Ga orbitals), and their balance with the MOF–water interactions.

Even inorganic materials, often considered as relatively rigid, can be driven to exhibit some level of flexibility under strong intermolecular interactions. This is the case of the inorganic framework zirconium tungstate ($ZrW_2O_8$), which shows "negative hydration expansion", i.e., a contraction of its lattice upon uptake of water. $ZrW_2O_8$ has been widely studied as a negative thermal expansion (NTE) material, meaning its volume shrinks when heated, over a wide range of temperature — a behavior for which it is known since the 1960s.[30] It was also known that the formation of the monohydrate $ZrW_2O_8 \cdot H_2O$ leads to a ~10% decrease in volume of the material's unit cell, although the microscopic explanation of this surprising hydration-induced shrinkage was not clear.[31,32] We have recently combined X-ray scattering measurements of the hydrated and dehydrated phases with Density Functional Theory (DFT) calculations of their atomic structures and their vibrations modes in order to explain this unusual phenomenon.[33] We identified a unique site for the strong physisorption of water, explaining the stoichiometric formation of the monohydrate, and the reversibility of adsorption.

Because there is a choice of three possible orientations for each water molecule, the hydrated state is heavily disordered, which is why its structure was not solved before. The concerted formation of new W–O bonds yields one-dimensional strings within the framework, obeying local connectivity rules but disordered at the larger scale. The entropically-favored macroscopic state for these self-avoiding strings is that of tangled spaghetti (see Figure 4). This was verified by actually creating mesoscale models of the disordered spaghetti-like states through a custom Monte Carlo algorithm, confirming that the generated models matched the signature X-ray pair distribution functions (PDF) obtained experimentally. Furthermore, we could establish an unexpected link, in that the concerted atomic displacements involved in the negative hydration expansion strongly implicate the vibrations modes that drive the negative thermal expansion. These rigid unit modes[34,35,36] feature correlated translation, rotation, and distortion of $ZrO_6$ and $WO_4$ polyhedra towards volume-reducing states. The fact that $ZrW_2O_8$ behaves as a metamaterials both under temperature changes and water adsorption offers an interesting perspective for other framework materials. It is possible that other materials, characterized and known for one specific behavior under stimulation, could be hiding other unusual properties — all linked in the same way to their structure, topology, and their constituting interactions.

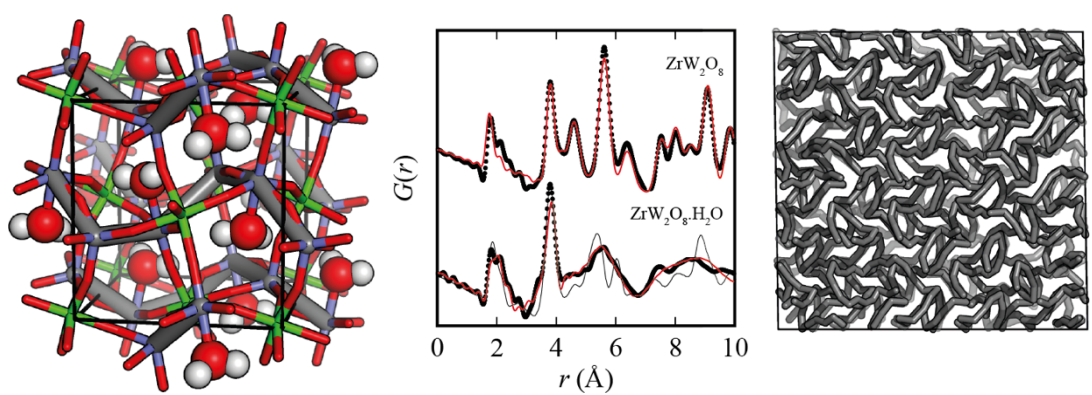

**Figure 4.** Left: Structure of $ZrW_2O_8$ upon hydration. Center: X-ray pair distribution functions of anhydrous and hydrated samples, compared with models (in red). Right: Model for a ZrW2O8 · H2O spaghetti configuration, where only W─O─W linkages are shown. Reproduced with permission from Ref. 33. Copyright 2018 the American Physical Society.

Another example is the study of $Zn_2(BDC-TM)_2(DABCO)$ by Burtch et al.[37] By combining *in situ* synchrotron diffraction, infrared spectroscopy and molecular modelling, they elucidated the water loading-dependent behavior of a flexible MOF. In this water-stable material, they could show that even as the crystallinity and porosity are maintained, important changes happen "under the hood", in the details of the structure due to water-induced bond rearrangements. Finally, we also note that while adsorption of water can trigger large-scale structural transitions, it can also be the cause of drastic changes in the properties of a framework. One such example is the reversible insulator-to-proton-conductor transition observed in a dense MOF, ((CH$_3$)$_2$NH$_2$)$_2$[Li$_2$Zr(C$_2$O$_4$)$_4$], where the adsorbed water molecules act as both proton source and proton carriers.[38]

**Reactivity in the presence of water**

Because water is a polar fluid, it can have strong and specific interactions with porous materials, through hydrogen bonding in particular, as described above. However, it is also possible that these interactions grow beyond the regime of physisorption, and that the water molecules are chemisorbed, i.e., form new chemical bonds. Metal–organic frameworks are particularly sensitive to this phenomenon, for several reasons. First, they often present coordinatively unsaturated metal sites (also called "open metal sites),[39,40] which are accessible for water to bond.[41,42] This, in turn, can drastically affect the properties of the MOF, either by poisoning the active metal sites,[43,44] or by increasing the framework's affinity for guest molecules through interactions with the adsorbed water.[45,46]

Secondly, chemisorption of water in MOFs occurs in some cases because strong metal–water interactions can disrupt the existing coordination interactions, leading to decoordination of organic linkers, possibly to a novel phase of the material, or even to the complete breakdown of the framework.[47,48] As a consequence, several MOFs are unstable in the presence of water[49,50] — including the archetypal MOF-5, featuring zinc–carboxylate coordination.[51] There is therefore an important drive to better understand what makes a material stable (or unstable) in the presence of water, including through molecular simulation, in order to improve the stability of existing materials — their performance in industrial applications being otherwise limited by a lack of stability at high temperature and in the presence of water. Because classical simulations cannot properly reproduce the strong intermolecular interactions involved, and the changes in bonding that may occur,[52] quantum chemistry-based methods are required.

Early studies often involved static DFT calculations[53,54] of water binding energies, its influence on the structure, and adsorption of other guest molecules. Later studies shifted the focus to a more dynamic picture of the water stability, using Born–Oppenheimer Molecular Dynamics (BOMD) based on DFT calculations of the interatomic forces.[55,56,57] These allow to quantify the impact of the level of hydration, and correctly reproduce a liquid environment — and not solely a few H$_2$O molecules in the gas phase. They can observe directly the hydrothermal instability of the framework, at a given temperature and water loading: the mechanism, kinetics, and energetics involved can be obtained.

However, in some cases, the process of hydrothermal breakdown is not fast enough to be observed spontaneously within the time scale of an *ab initio* MD simulation. In such cases, the binding of water molecules and displacement of the existing framework coordination is an activated event, which can lead to instability at the macroscopic timescale, but still be too slow to be observed within the typical time of such simulations, from tens to a hundred of picoseconds. This was the case in a study of the hydrothermal breakdown of MIL-53(Ga) where we used the metadynamics free energy technique,[58] in combination with *ab initio* MD, in order to elucidate the mechanism of instability. We could confirm that the weak point of the structure is the bond between the metal center and the organic linker, and highlight how the presence of water lowers the activation free energy for the breaking of the metal–ligand coordination.[59] This is quantitatively measured by free energy profiles, plotted as a function of one or two reaction coordinates of choice for the system, as depicted in Figure 5.

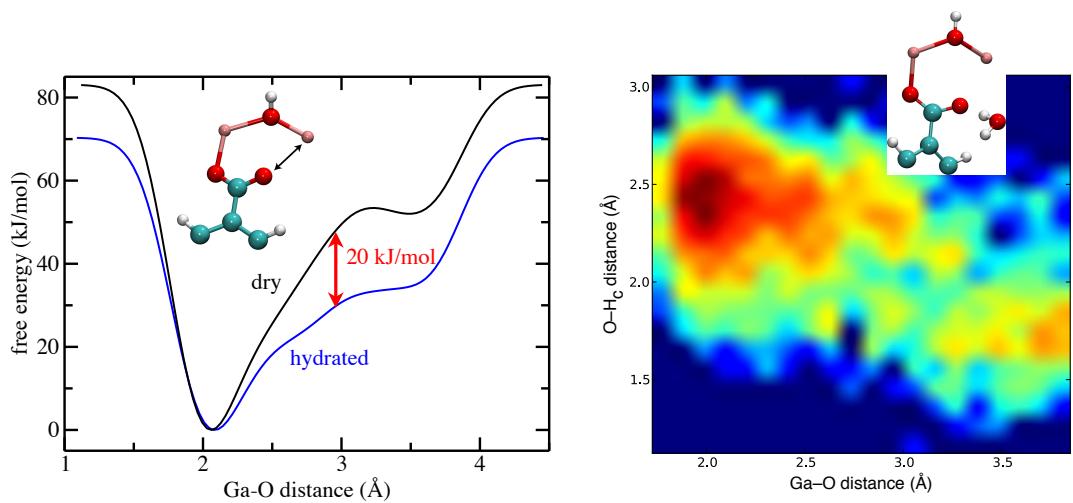

**Figure 5.** Left: Free energy profiles of Ga−O bond breaking for dry (black) and hydrated (blue) MIL-53(Ga) at 650 K, obtained from metadynamics *ab initio* MD simulations. Right: two-dimensional free energy map for MIL-53(Ga) in the presence of water, as a function of Ga–O and O–H distances. Reproduced with permission from Ref. 59. Copyright 2015 American Chemical Society.

Another example of a chemically heterogeneous surface, near which the behavior of water is complex, is the case of graphene oxide.[60] It is a 2D graphene-based material whose properties strongly depend on the oxygen-bearing functional groups present at its surface: alcohol, epoxide, ketones, and carboxylic acid, etc.[61,62] When it is present in liquid water (e.g., in graphene oxide membranes), it exists as a stacking of layers with a distribution of interlayer distances of the order of one nanometer, with confined water in this interlayer space. While graphene oxide has potential applications for water purification and desalination,[63] the nature of the chemical interactions between graphene oxide and liquid water remain poorly understood, and are difficult to probe experimentally because of the nonregular distribution of the oxygen groups. In order to shed light into the behavior of the graphene oxide/water interface, we have recently performed first-principles molecular dynamics simulations on a series of graphene oxide models with identical chemical composition, but different distributions of the functional groups.[64] Our work shows that hydroxyl and epoxide groups preferentially gather in oxygen-rich zones, leaving uncovered areas ("ribbons") of pristine graphene. Starting from a neutral graphene oxide sheet and neat water, simulations also show that reactivity can occur at the interface, leading to a negative electric charge on the graphene oxide sheet by deprotonation (and, in some cases, dehydration;

both event are depicted on Figure 6). It is important to note that such behavior, because it involves the breaking and formation of bonds, could not be studied by classical simulation methods, and requires CPU-expensive first-principles methods to include both a description at the quantum chemical level, and a statistical (and dynamical) sampling of the configurations of this complex system in the liquid phase. It is therefore an interesting demonstration of the phenomena that can be probed with *ab initio* molecular dynamics.

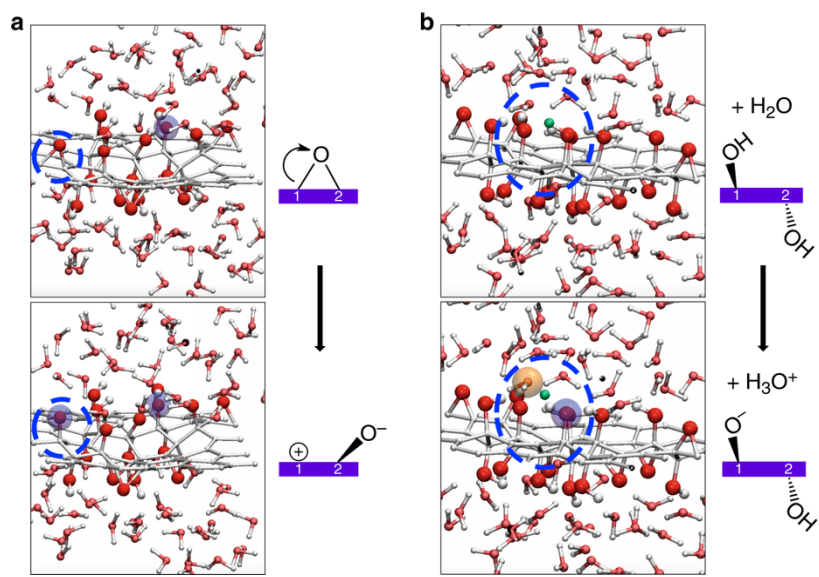

**Figure 6.** Snapshots taken from first-principles molecular dynamics simulations of graphene oxide in liquid water. Reproduced with permission from Ref. 64. Copyright The Authors, some rights reserved. Distributed under a Creative Commons Attribution License 4.0 (CC BY) https://creativecommons.org/licenses/by/4.0/

**Perspectives**

The above sections have presented several examples of unusual properties of water confined in heterogeneous nanopores, and of surprising behavior arising from the interaction between the water and the soft porous crystals. In particular, the synergy between advanced experimental characterization methods and computational modelling was highlighted, as a way to provide a broad picture of the systems and phenomena — covering the questions of structure, dynamics, response to external stimulation, and chemical reactivity. One of the recent trends observed in the literature is to use computational methodologies and in situ characterization to address systems that are more and more complex, both in their chemistry and the geometry and topology of their pore network. In particular, the focus is shifting slowly from being purely on crystalline systems, to deal with disordered and amorphous phases: MOFs are now considered for applications as amorphous materials,[65] gels,[66] or even liquids — and composites thereof![67] Modelling methods need to be expanded to be able to deal with these more complex systems, extending the length scales that can be addressed. Because of the high computational cost and scaling of quantum methods, one of the ways to achieve this is the development of high-accuracy force fields based on quantum chemistry reference data.[68,69] Another one is to mesh together ab initio methods, classical methods, and coarse-grained (or mesoscale) simulation of porous materials[70] in order to bridge these scales into a truly multi-scale modelling strategy.

Finally, another perspective is the study of more complex adsorbed systems: metal–organic frameworks, and other nanoporous materials, are being proposed for applications such as liquid separation, where there is a dire need of understanding the influence of confinement on liquid mixtures and liquid/liquid phase diagrams. Other recent studies have focused on the properties of aqueous solutions of high concentration (such as LiCl up to 20 mol L$^{-1}$), confined in flexible nanoporous materials,[71] demonstrating that they can have specific properties different from the pure water,[72] and not entirely explained by osmotic effects.[73] There, the strong ordering of the confined electrolyte competes with the structural flexibility of the framework to create an entirely new behavior for the {host, guest} system, and it is important to develop new methods to better understand such processes,[74] both for their fundamental importance but also for their proposed applications in energy storage and dissipation systems.[75]

**Notes**

The authors declare no competing financial interest.

**Biography**

François-Xavier Coudert is a Senior Researcher (*directeur de recherche*) at the French National Centre for Scientific Research (CNRS), where his group applies computational chemistry methods at various scales to investigate the physical and chemical properties of nanoporous materials, and in particular stimuli-responsive materials with anomalous behavior. He obtained his PhD from the University Paris-Sud (France) in 2007, for work on the properties of water and solvated electrons confined in zeolite nanopores. He worked as post-doctoral researcher at University College London (UK) on the growth of metal–organic frameworks on surfaces, before joining CNRS in 2008.

**Acknowledgments**

The very nature of research is that of a collaborative enterprise, and I thank Alain Fuchs and Anne Boutin for our continuing collaboration on several of the topics discussed here, many great scientific discussions and fun projects. I also thank Camille Noûs and Ike Antkare for their influence on this work.